\documentclass{emulateapj}

\usepackage{graphicx}
\usepackage{subfigure}
\shorttitle{{\it WMAP} and {\it EGRET} Cross-Correlation}
\shortauthors{X. Liu and S. N. Zhang}

\begin{document}


\title{Cross-Correlation analysis of {\it WMAP} and {\it EGRET} in Wavelet Space}


\author{Xin Liu\altaffilmark{1} and Shuang Nan Zhang\altaffilmark{1,2,3,4}}
\altaffiltext{1}{Physics Department and Tsinghua Center for
Astrophysics, Tsinghua University, Beijing 100084, China;
liux00@mails.tsinghua.edu.cn, zhangsn@tsinghua.edu.cn.}
\altaffiltext{2}{Key Laboratory of Particle Astrophysics,
Institute of High Energy Physics, Chinese Academy of Sciences,
P.O. Box 918-3, Beijing 100039, China}
\altaffiltext{3}{Physics Department, University of Alabama in Huntsville, Huntsville,
AL 35899, USA}
\altaffiltext{4}{Space Science Laboratory, NASA Marshall Space Flight Center, SD50,
Huntsville, AL 35812, USA}
%

\begin{abstract}
We cross correlate the {\it Wilkinson Microwave Anisotropy Probe}
({\it WMAP}) first year data and the diffuse gamma-ray intensity
maps from the {\it Energetic Gamma Ray Experiment Telescope} ({\it
EGRET}) using spherical wavelet approaches. Correlations at
$99.7\%$ significance level have been detected, at scales around
$15^{\circ}$ in the {\it WMAP} foreground cleaned W-band and
Q-band maps, based on data from regions that are outside the most
conservative {\it WMAP} foreground mask; no significant
correlation is found with the Tegmark cleaned map. The detected
correlation is most likely of Galactic origin, and thus can help
us probing the origins of possible Galactic foreground residuals
and ultimately removing them from measured microwave sky maps.
\end{abstract}





\section{Introduction}
The study of the anisotropies of the cosmic microwave background
(CMB) is a powerful tool in cosmology. Results from the Wilkinson
Microwave Anisotropy Probe ({\it WMAP}) provide us with the
angular power spectrum and its cosmological implications (Bennett
{\it et al.} 2003a; Spergel {\it et al.} 2003; Page {\it et al.}
2003). However unwanted signals due to foregrounds would
contaminate any intrinsic signals, most importantly on large
angular scales (Tegmark \& Efstathiou 1996; de Oliveira-Costa {\it
et al.} 2003).

The {\it WMAP} team chose the CMB-dominated bands (two Q-band maps
at 40.7 GHz, two V-band maps at 60.8 GHz and four W-band maps at
93.5 GHz) and combined them to give a signal-to-noise ratio
enhanced map. Despite that the maps at selected frequencies are
dominated by CMB, Galactic foregrounds as well as extragalactic
point sources all contribute significantly to the map,
where the {\it WMAP} team performed a foreground template fit
(thermal dust from Finkbeiner {\it et al.} 1999; free-free from
Finkbeiner 2003 and Schlegel {\it et al.} 1998; synchrotron from
Haslam {\it et al.} 1982) to avoid the Galactic emissions and
separated them from the underlying CMB signal according to the
frequency-dependence property of foregrounds (Bennett {\it et al.}
2003b).
Since the foreground fluctuations depend on the multipole moment
$l$ as well (Bouchet {\it et al.} 1995), $l$-dependent statistical
weights have been applied in Tegmark {\it et al.} (2003) where an
independent foreground analysis was made. Several works of {\it
WMAP} non-Gaussianity detection found residuals or the systematic
effect correction as the sources (Chiang {\it et al.} 2003; Chiang
\& Naselsky 2004; Naselsky, Doroshkevich \& Verkhodanov 2003,
2004; Eriksen {\it et al.} 2004; Hansen {\it et al.} 2004; Liu \&
Zhang 2005).

Wibig and Wolfendale (2005) have shown evidence that the Tegmark
cleaned map contains residual foregrounds possibly induced by
cosmic rays, where the Energetic Gamma Ray Experiment Telescope
({\it EGRET}) diffuse gamma-ray intensity map (Hunter {\it et al.}
1997) was adopted as the Galactic tracer. Diffuse Galactic
gamma-ray emission is supposedly produced in interactions of
Galactic cosmic rays with the interstellar gas and radiation
field, and thus provides us with an indirect measurement of cosmic
rays in various locations in the Galaxy \footnote{See Bertsch {\it
et al.} 1993 and Hunter {\it et al.} 1997 for three-dimensional
modeling.} In this paper we cross-correlate the {\it WMAP}
first-year data and the {\it EGRET} maps, which are based on more
data and more complete point-source subtraction (Cillis \& Hartman
2005), in wavelet space to probe the origins of potential residual
foregrounds with characteristic scales. Both the {\it WMAP}
combined foreground cleaned maps (Bennett {\it et al.} 2003b) and
the Tegmark cleaned map (Tegmark {\it et al.} 2003) have been
used.

\section{Cross-correlation in wavelet space}
A measure of the correspondence of two data sets on the sphere is
the angular cross-correlation function $CCF(\theta)$, which
represents how the two measures of the sky separated by an angle
$\theta$ are correlated. Previous works have performed the cross
correlation of the CMB data and the nearby galaxy density tracers
in search for the Integrated Sachs-Wolfe effect (ISW) (Boughn \&
Crittenden 2002, 2004; Nolta {\it et al.} 2004; Fosalba \&
Gazta\~{n}aga 2004; Fosalba {\it et al.} 2004; Afshordi {\it et
al.} 2004; Vielva {\it et al.} 2004b). As being performed in
Vielva {\it et al.} (2004b), cross-correlation studies can also be
made in wavelet space.

Wavelet approach is very useful for detecting signals with a
characteristic scale that a most optimal detection can be made by
filtering the data at a given scale, thus structures at that scale
are amplified. It has been adopted in the CMB-related analysis for
non-Gaussianity studies (Hobson {\it et al.} 1999; Barreiro {\it
et al.} 2000; Aghanim {\it et al.} 2003; Cay{\'o}n {\it et al.}
2001, 2003; Mart{\'\i}nez-Gonz{\'a}lez {\it et al.} 2002;
Mukherjee \& Wang, 2004; Vielva {\it et al.} 2004a; McEwen {\it et
al.} 2004; Liu \& Zhang 2005).
In wavelet space, the cross-correlation covariance at a given
scale $a$ is defined as (Vielva {\it et al.} 2004b):
\begin{equation}\label{cov}
Cov_{_{{\tiny{W-E}}}} (a)=
\frac{1}{N_{a}}\sum_{\vec{p}}{\omega}_{_{{\tiny{CMB}}}}(a,
\vec{p}) {\omega}_{_{{\tiny{EGRET}}}}(a, \vec{p}),
\end{equation}
where ${\omega}_{_{{\tiny{CMB}}}}(a, \vec{p})$ and
${\omega}_{_{{\tiny{EGRET}}}}(a, \vec{p})$ are the wavelet
coefficients of the {\it WMAP} and {\it EGRET} data at position
$\vec{p}$ on the sky map, and the sum $\sum_{\vec{p}}$ is extended
over all the pixels ($N_{a}$) that are not masked by the Galactic
mask ``Kp0". Equation (1) gives the auto-correlation covariance
(ACC) when the two data sets are the same. To make the analysis
more robust and less sensitive to any discrepancies between CMB
data and simulations, we use the dimensionless and normalized
cross-correlation coefficient $C_{_{{\tiny{W-E}}}}(\theta)$ as the
test statistics, which is given by
$C_{_{{\tiny{W-E}}}}(\theta)=Cov_{_{{\tiny{W-E}}}}(\theta)/(\sigma_{_{{\tiny{W}}}}\sigma_{_{{\tiny{E}}}})$,
where $\sigma_{_{{\tiny{W}}}}^2=ACC_{_{{\tiny{W}}}}$ and
$\sigma_{_{{\tiny{E}}}}^2=ACC_{_{{\tiny{E}}}}$ are the {\it WMAP}
and {\it EGRET} auto-correlation covariances respectively.

The wavelet coefficients are obtained by convolving the map with a
certain spherical wavelet basis at a given scale:
\begin{equation}
{\omega}_{_{{\tiny{D}}}}(a, \vec{p}) = \int d\Omega' \, D(\vec{p}
+ \vec{p'})
    \Psi_S(\theta', a),
\end{equation}
where $\Psi_S(\theta', a)$ is the spherical wavelet basis and
$D(\vec{p} + \vec{p'})$ is the data set to be analyzed. The
spherical wavelet can be obtained from the Euclidean Wavelet
counterpart following the stereographic projection suggested by
Antoine \& Vanderheynst (1998). Mart{\'\i}nez-Gonz\'alez {\it et
al.} (2002) has described the projection for the Spherical Mexican
Hat Wavelet (SMHW) as well as its properties, whereas the
Spherical Morlet Wavelet (SMW) has been applied for
non-Gaussianity detection in the {\it WMAP} data (McEwen {\it et
al.} 2004, Liu \& Zhang 2005). In this correlation study we have
adopted both wavelets, of which the SMW seems to be more sensitive
than the SMHW. We only present results from the SMW analysis here,
even though similar correlations are also found when applying the
SMHW.

The preprocessing pipelines for the {\it WMAP} CMB data and Monte
Carlo simulations are the same as in Liu \& Zhang (2005), whereas
the gamma-ray intensity maps are degraded to the same resolution
with the CMB data to be cross-correlated, all in equi-angular
spherical grid pixalisation. The preprocessed maps are shown in
Fig.~1. 10,000 simulations of the {\it WMAP} data have been
applied, following the cosmological model given by the Table 1 of
Spergel {\it et al.} (2003), to obtain the significance levels of
any correlation detection. We obtain these levels in a robust way
by taking into account the probability distribution of the wavelet
cross-correlation coefficient at each scale. Although the
map-making algorithm presented by Cillis and Hartman (2005)
possibly produced some systematic effects in the {\it EGRET} data,
these can be calibrated out by simulations.


\begin{figure}\label{prepro}
    \centering\subfigure
    {\includegraphics[scale=0.33]{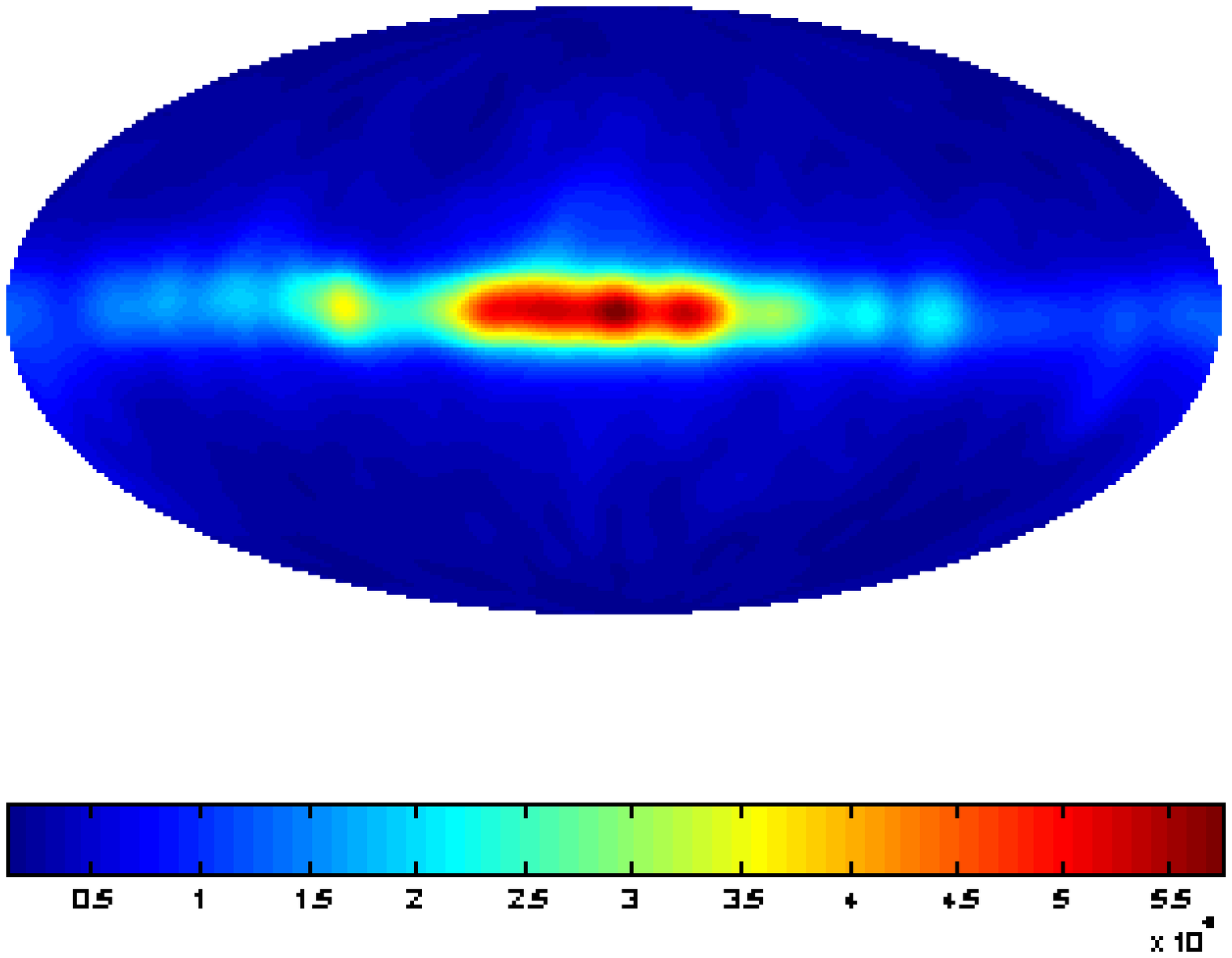}}
    \centering\subfigure
    {\includegraphics[scale=0.33]{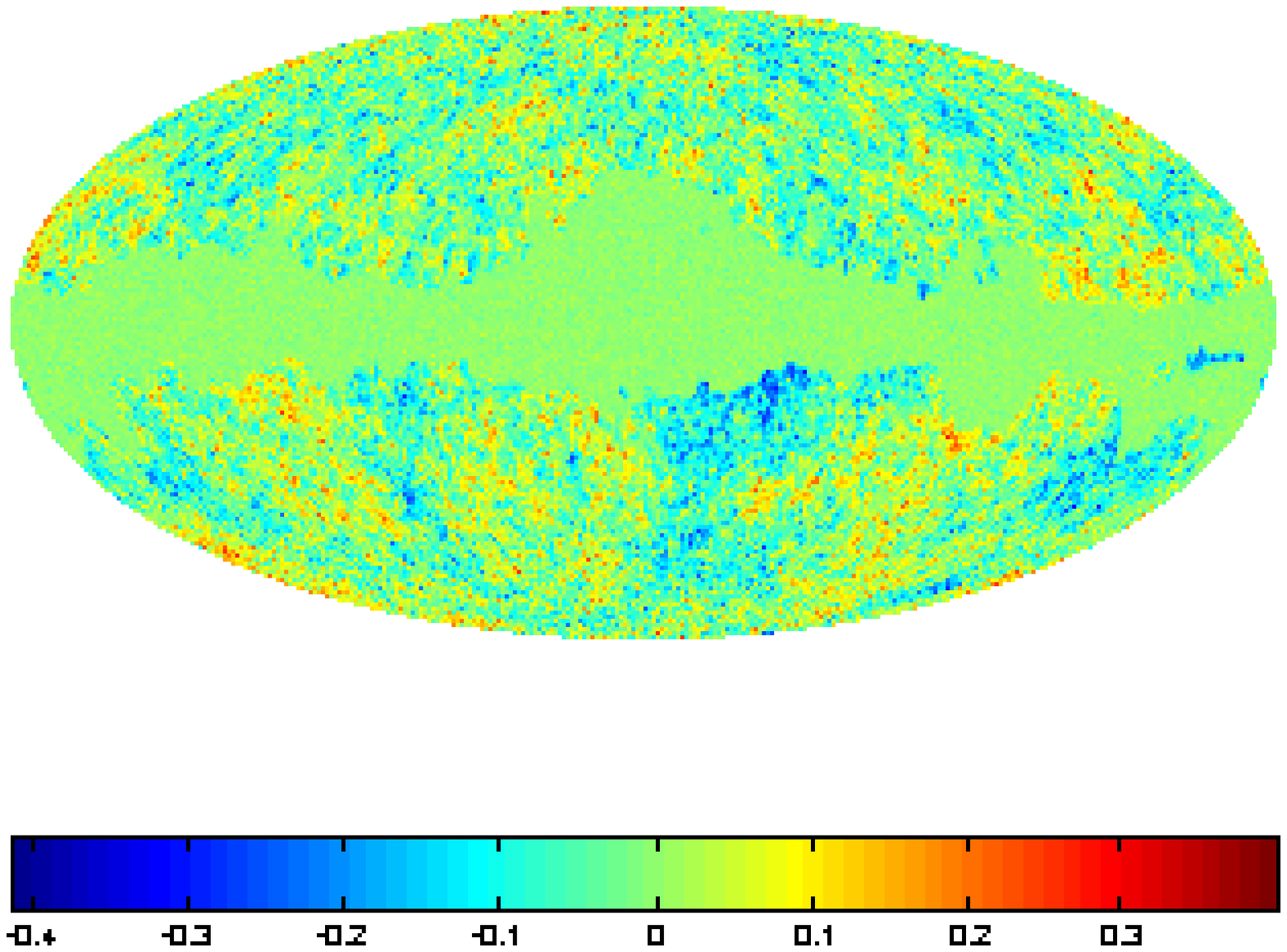}}
    \centering\subfigure
    {\includegraphics[scale=0.33]{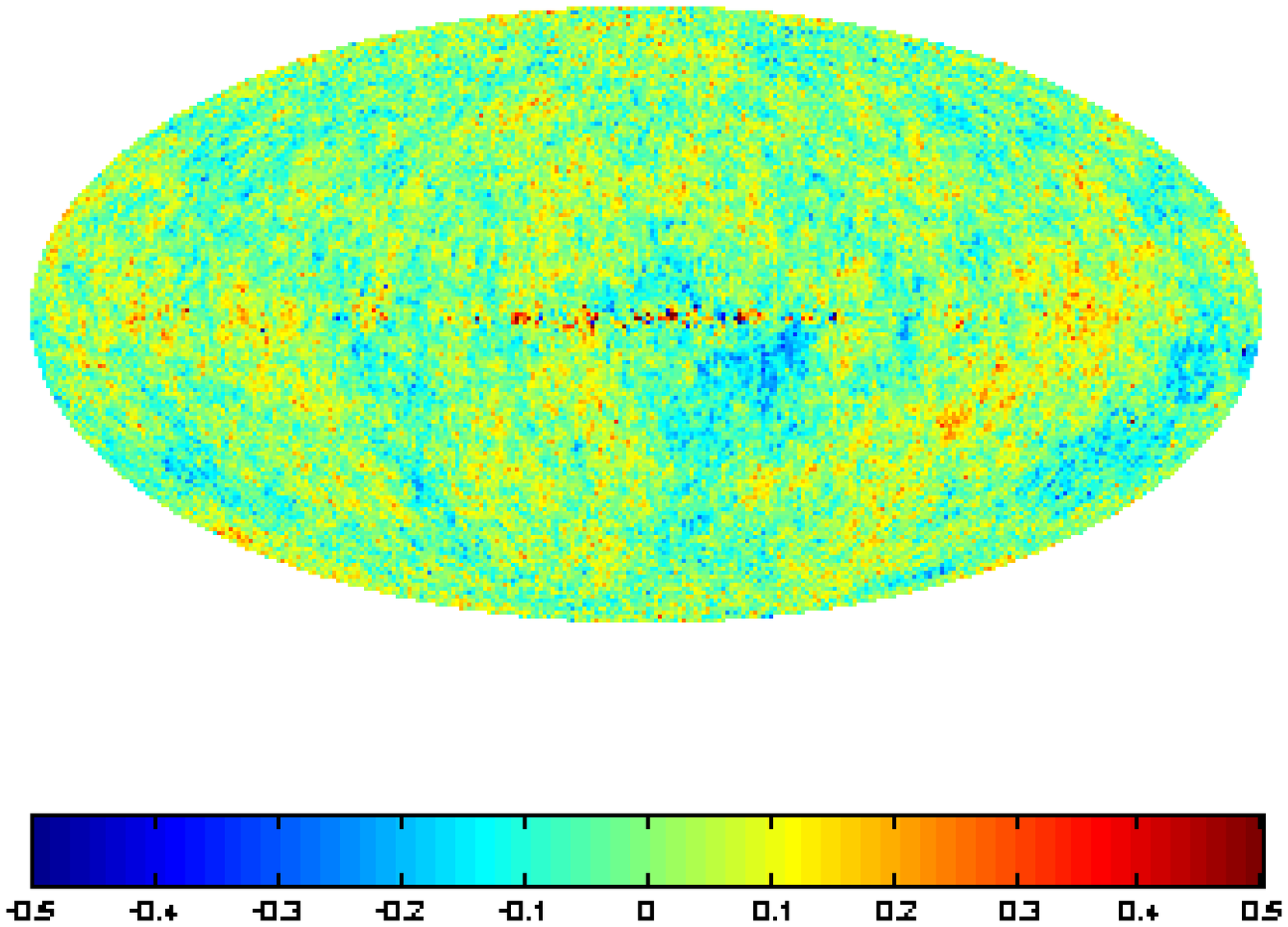}}
    \caption{Preprocessed {\it EGRET} diffuse gamma-ray intensity map ($>1$ GeV)
    (Cillis and Hartman 2005) and CMB anisotropy maps from the {\it WMAP} map
    (Bennett {\it et al.} 2003a) and the Tegmark cleaned map
    (Tegmark {\it et al.} 2003) to be analyzed in the cross-correlation study.
    Note that we have analyzed the EGRET diffuse gamma-ray intensity maps
     of different energies separately, i.e., $30-100$ MeV, $100-300$ MeV,
     $300-1000$ MeV and $>1$ GeV, where only the $>1$ GeV map is shown here.
     All maps are plotted in Galactic coordinates with
     the Galactic center $(l,b)=(0,0)$ in the middle and Galactic
     longitude $l$ increasing to the left.}
\end{figure}

\section{Results}
Cross correlation with the {\it EGRET} maps in wavelet space has
been performed to both of the Q-V-W combined {\it WMAP} map and
the Tegmark cleaned map to detect foreground residuals (here and
throughout, all {\it WMAP} CMB maps used are foreground removed
maps). Also we have performed the analysis to: (a) the {\it WMAP}
map in each band separately, i.e. Q, V, and W; (b) systematic beam
and noise maps from subtracting maps of the receivers at the same
frequency; and (c) a foreground map almost free of CMB, which is
made by subtracting the two receivers of Q band and the two
receivers of V band from the four receivers of W band (Vielva {\it
et al.} 2004a).

For every CMB related map, we have performed the analysis using
{\it EGRET} maps in different energy ranges separately, i.e.
$30-100$ MeV, $100-300$ MeV, $300-1000$ MeV and $>1$ GeV, and only
the results from the $>1$ GeV map are shown here because: (a) it
has been stated by Cillis \& Hartman (2005) that the $>1$ GeV
intensity map has the best statistical accuracy compared with
others from lower energy ranges, where possible correlations can
be smeared out by statistical uncertainty fluctuations; and (b)
the analysis with different gamma-ray energy ranges stated above
generally give similar correlations and the significance level
seems to increase a little with the energy of the adopted
gamma-ray data. Note this variance of correlation significance
does not conflict with the power-law energy spectrum, since the
cross-correlation coefficient is a normalized quantity.

The cross-correlation coefficients $C_{_{{\tiny{W-E}}}} (\theta)$
are illustrated in Figs. 2(a) and 2(b), where the corresponding
size $\theta$ on the sky can be obtained from equation (3) in Liu
\& Zhang (2005); we have analyzed scales from $1^{\circ}$ to
$90^{\circ}$, where we are only concerned about results at scales
$> 4^{\circ}$ due to the angular resolution limit in the {\it
EGRET} data. We have shown the ACCs (normalized by the map
dispersion $\sigma_0$ in the real space) of the {\it WMAP}
combined map, the Tegmark cleaned map, the foreground-component
map and the diffuse gamma-ray intensity map in Fig. 2(c), which
indicate that the general patterns of $C_{_{{\tiny{W-E}}}}
(\theta)$ in Figs. 2(a) and 2(b) are caused by the convolution of
the gamma-ray data and the wavelets, not by systematical
artifacts. Note at all the concerned scales the foreground map
correlates more significantly with the {\it EGRET} map, consistent
with the assumption that most of the foregrounds have been removed
from the CMB maps. Results from analyzing the systematic beam and
noise maps are not shown since their correlation coefficients are
all consistent with zero values within statistical fluctuations.

For the {\it WMAP} CMB maps, correlations at $99.7\%$ significant
level are detected at scales from about $14^{\circ}$ to
$16^{\circ}$, in the W-band and Q-band maps; less significant
correlations ($96.8\%$) are found at larger scales, from about
$43^{\circ}$ to $48^{\circ}$, only in the W-band map. In sum, the
W-band and Q-band maps exhibit more significant correlations
compared with those from the V-band map, at all the concerned
scales. This frequency dependence feature can be easily understood
according to Fig. 10(a) in Bennett {\it et al.} (2003b), showing
evidence that the detected correlations are caused by residual
foreground signals. In order to test whether the detected signal
has a Galactic origin, we have shown in Fig. 2(d) the
cross-correlation coefficients as a function of the galactic
latitude around $\theta=15^{\circ}$. Note that the sum of the
cross-correlation coefficients at all the latitudes in Fig. 2(d)
corresponds to the $C_{_{{\tiny{W-E}}}} (15^{\circ})$ in Figs.
2(a). Here the oscillating pattern is given by the SMW, and we
have tested that the outline profile is caused by the {\it EGRET}
intensity maps. Tests at other scales also present similar
profiles. The results show some consistency with the co-sec law
for the Galactic components, and also present a north-south
asymmetry where the correlation is stronger in the southern
hemisphere.

Analysis of the Tegmark cleaned map does not show any significant
correlation at all the concerned scales, as an evidence that it is
``cleaner" than the {\it WMAP} combined map.

\begin{figure}
    \centering\subfigure
    {\includegraphics[scale=0.32]{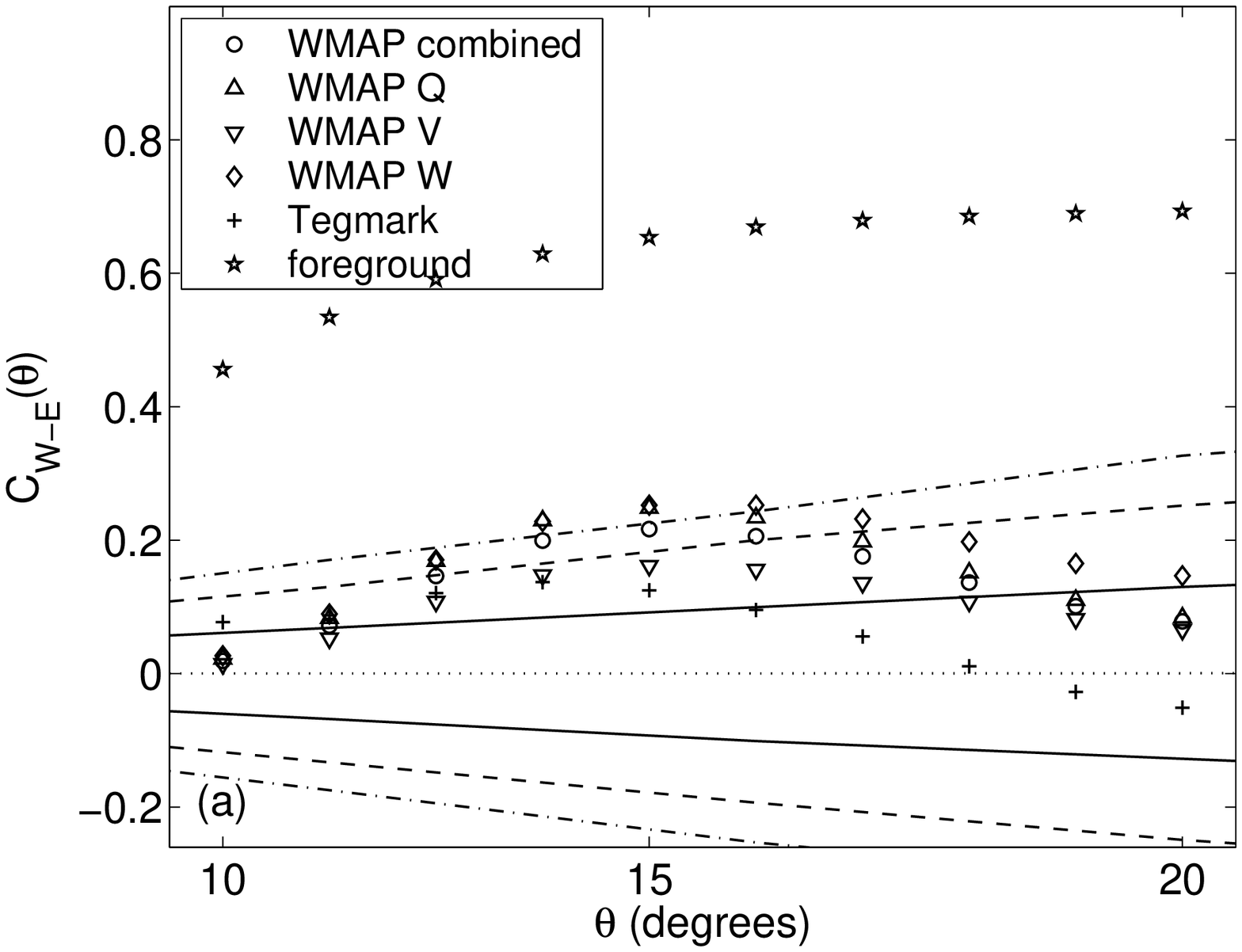}}
    \centering\subfigure
    {\includegraphics[scale=0.32]{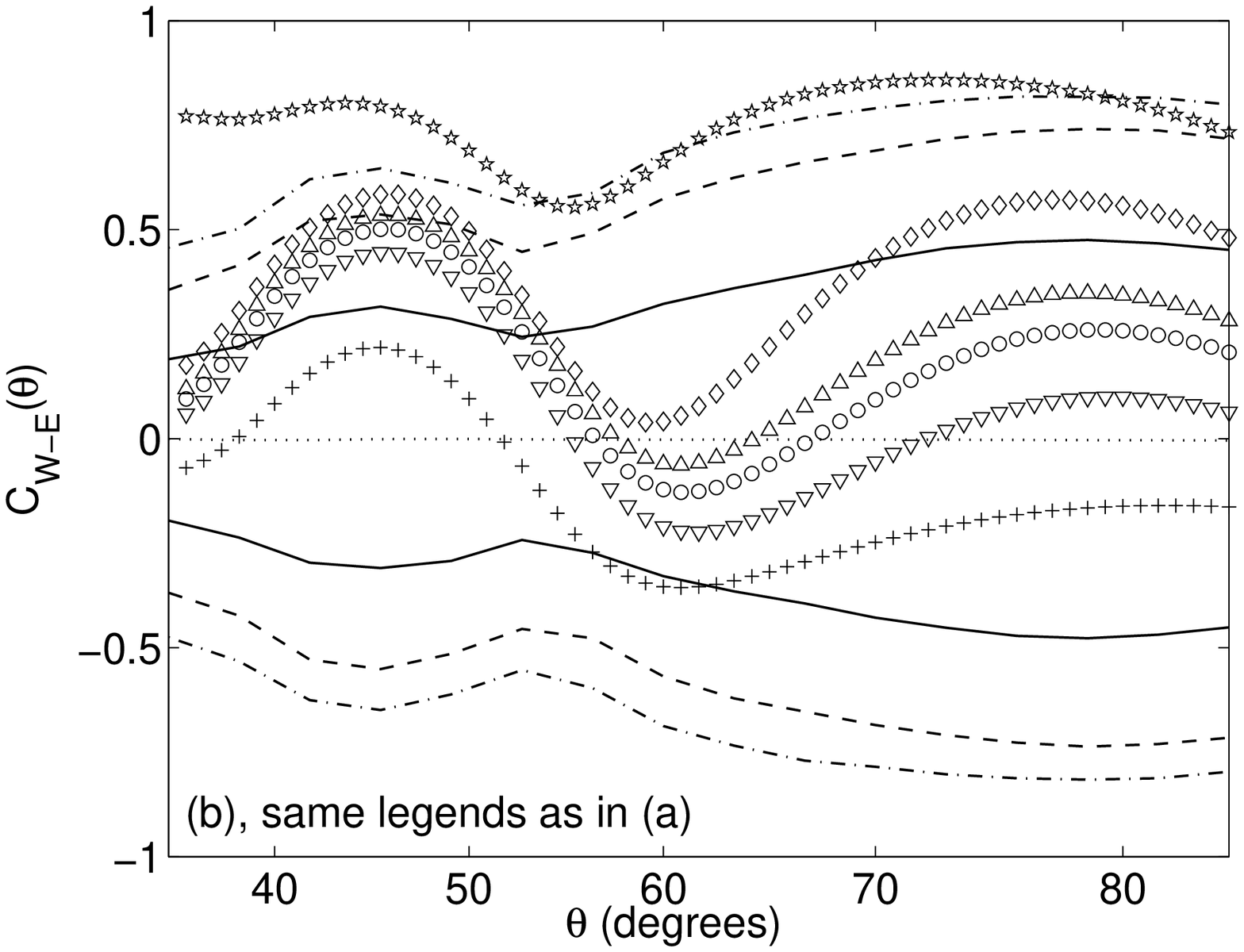}}
    \centering\subfigure
    {\includegraphics[scale=0.32]{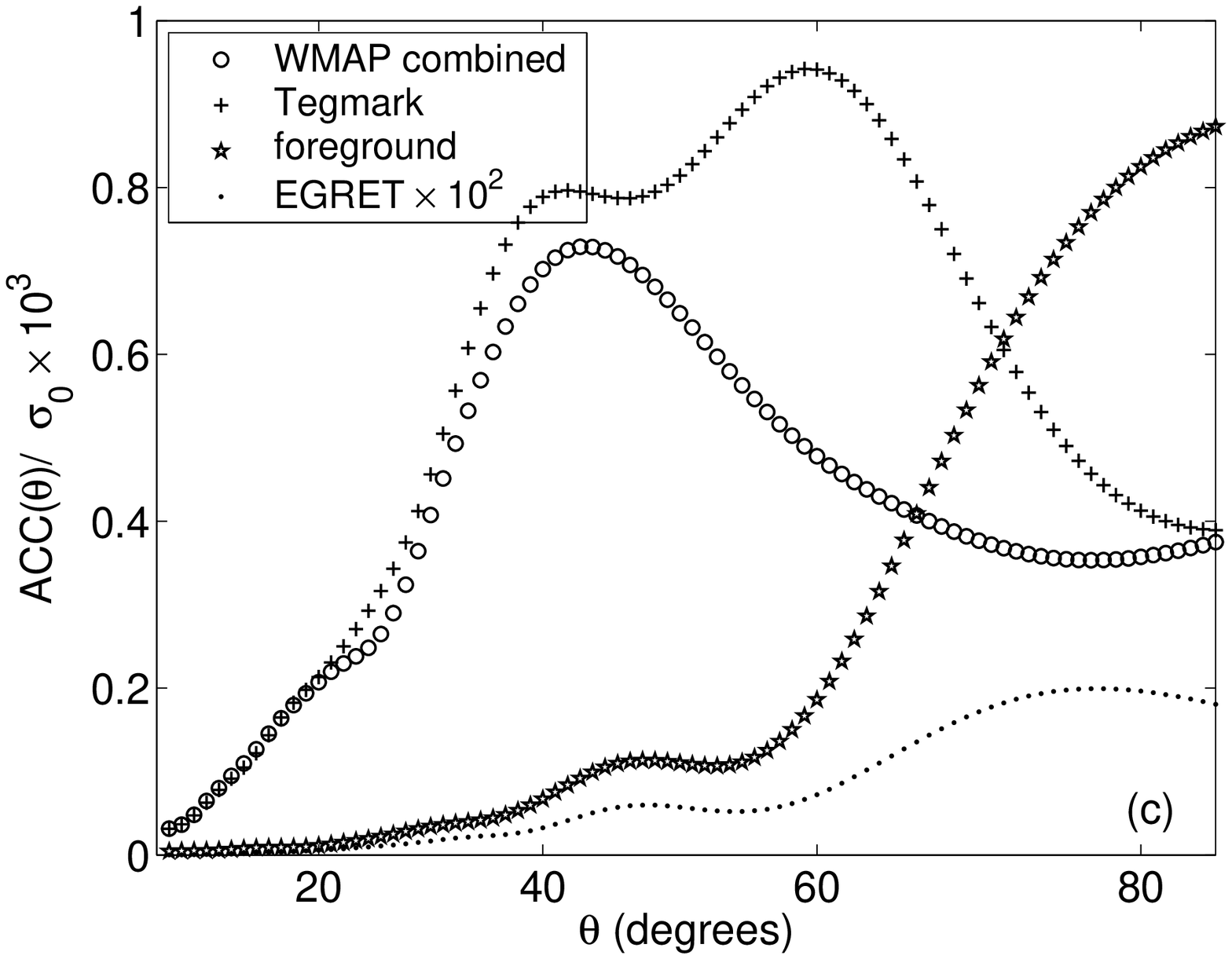}}
    \centering\subfigure
    {\includegraphics[scale=0.32]{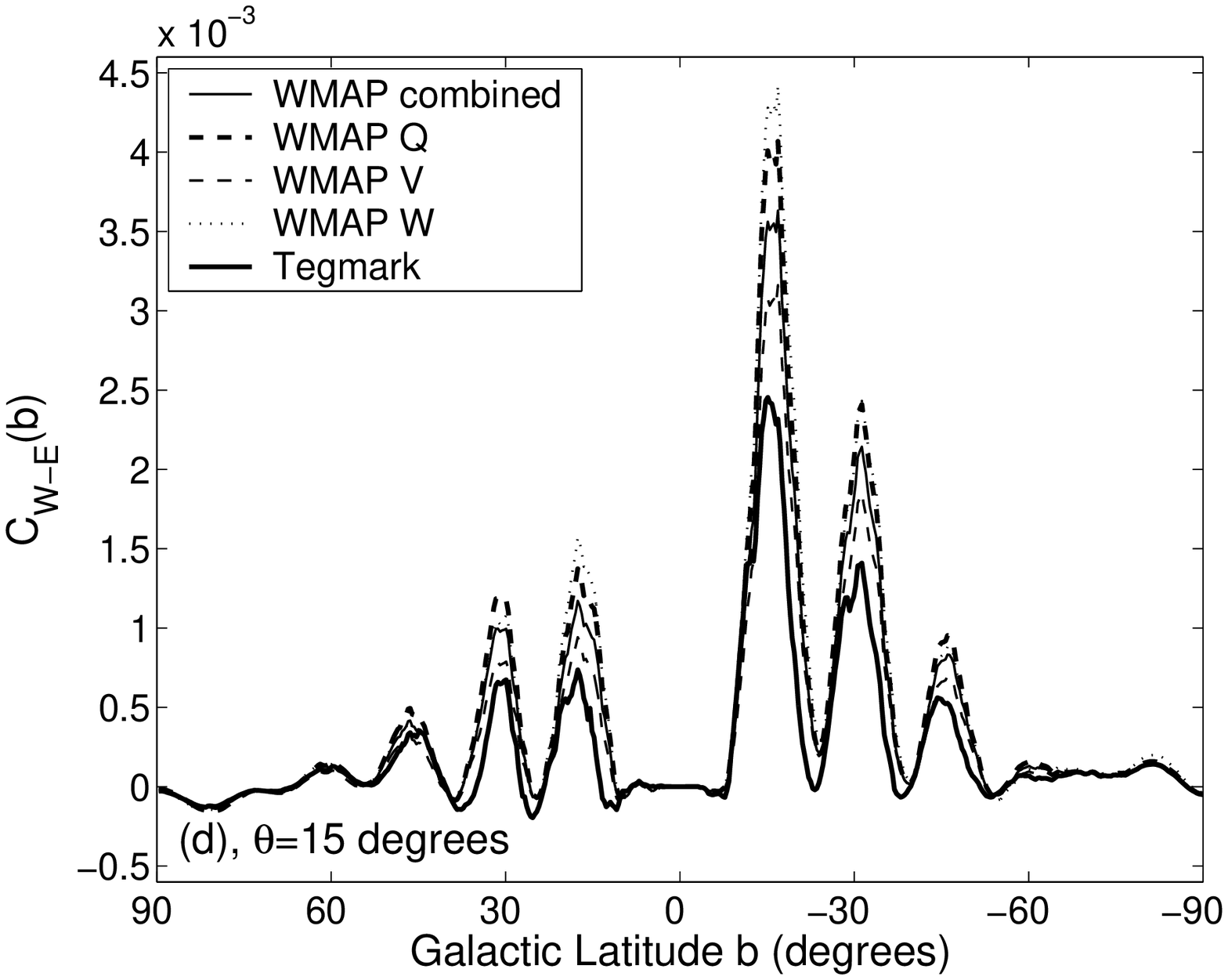}}
    \caption{(a) and (b): cross-correlation coefficients $C_{_{{\tiny{W-E}}}} (\theta)$
    of the {\it WMAP} and {\it EGRET} data in wavelet space. The acceptance intervals at the 68\% (solid), 95\% (dashed)
     and 99\% (dash-dotted) significance levels given by 10,000 Monte Carlo
    simulations are shown, respectively. The error bars are too small compared to the coefficient values
    and can be ignored in the figures; (c): normalized auto-correlation
    covariances, representing that the general
    patterns in (a) and (b) are caused by the convolution of the gamma-ray data and
    the wavelet basis, not by systematical artifacts; (d):
    cross-correlation coefficient as a function of the galactic
    latitude around $\theta=15^{\circ}$, testing the Galactic origin of the detected correlations.}
\end{figure}

\section{Conclusions and Discussions}

We have performed cross-correlation analysis of the {\it WMAP}
first-year data with the {\it EGRET} diffuse gamma-ray intensity
maps in wavelet space, finding correlations in the {\it WMAP}
foreground cleaned maps, based on regions that are outside the
most conservative {\it WMAP} foreground mask. Analysis of the {\it
WMAP} W-band and Q-band maps exhibits correlation at $99.7\%$
significant level at scales around $15^{\circ}$ in the sky, and at
scales around $45^{\circ}$ less significant correlation ($96.8\%$)
is found only in the {\it WMAP} W-band map. The Tegmark cleaned
map seems to be compatible with pure CMB simulations at all the
concerned scales.

These cross-correlation signals are not caused by systematic beams
nor noise because: (a) the analysis of the systematic beam and
noise map show almost zero correlation results; and (b) the
correlations are detected at the scales where noise and beam
effects can be ignored (Tegmark {\it et al.} 2003). We thus
conclude that these cross-correlation signals are most probably
caused by foreground residuals because: (a) correlations from the
Q and W bands are more significant than those from the V band,
consistent with the frequency dependence nature of foreground; (b)
the correlation coefficients from the CMB maps present similar
patterns with those from the foreground map, whereas random
simulations do not show correlations at the the detected level;
(c) the Tegmark cleaned map, which has been commonly believed to
be cleaner than the {\it WMAP} combined map in terms of the
foreground removal, shows no significant correlation with the {\it
EGRET} map in the analysis; and (d) the detected cross-correlation
coefficient as a function of the galactic latitude appears to be
consistent with the co-sec law as an evidence for the Galactic
origin. It is possible that these foreground residuals may be
induced by cosmic rays, since the Galactic diffuse gamma-ray
emission is supposedly produced in interactions of cosmic rays
with gas and ambient photon fields and thus can be an indicator of
cosmic rays in various locations in the Galaxy.

The detected foreground residuals can be located in the
coefficient map at a certain scale. It is therefore worthwhile to
make correlation study with cosmic ray's spatial and spectral
distributions in detail. Note that the diffuse gamma-ray emission
has not been well understood that at energies $>1$ GeV the
observed intensity in inner Galaxy displays a GeV excess at a
level of $60\%$ compared with predictions (Hunter {\it et al.}
1997), which has been interpreted in several models (Strong {\it
et al.} 2000; de Boer 2005). Further cross-correlation work must
be done to fully understand the nature of foreground residuals and
finally remove them from the CMB maps completely, in order to
minimize the impacts of foreground residuals to the cosmological
studies of CMB.


\noindent{\bf Acknowledgement: } We are very grateful to the
anonymous referees whose extremely detailed and insightful
comments and suggestions allowed us to clarify several issues and
improve the readability of the paper.
This research has made use of the Astrophysical Integrated
Research Environment (AIRE) which is operated by the Center for
Astrophysics, Tsinghua University.
We acknowledge the use of LAMBDA, support for which is provided by
the NASA Office of Space Science, the YAWtb
(http://www.fyma.ucl.ac.be/projects/yawtb) toolbox developed by A.
Coron, L. Jacques, A. Rivoldini and P. Vandergheynst, the software
packages HEALPix (http://www.eso.org/science/healpix), developed
by K.M. Gorski, E. F. Hivon, B. D. Wandelt, J. Banday, F. K.
Hansen and M. Barthelmann., and CMBFAST (http://www.cmbfast.org),
developed by U. Seljak and M. Zaldarriaga.
This study is supported in part by the Special Funds for Major
State Basic Research Projects and by the National Natural Science
Foundation of China. SNZ also acknowledges partial funding support
from the US NASA Long Term Space Astrophysics program.

\end{document}